\documentclass[a4paper]{jpconf}
\usepackage{graphicx}
\usepackage{iopams}
\usepackage{amssymb,latexsym}
\usepackage{amstext,amsthm}
\usepackage{amscd}
\usepackage{amsgen,amsfonts,amsbsy}
\theoremstyle{plain}

\theoremstyle{definition}

\newcommand{\Z}{\mathbb{Z}}
\newcommand{\N}{\mathbb{N}}

\newcommand{\C}{\mathbb{C}}
\newcommand{\set}[2]{\{#1|\ #2\}}
\newcommand{\sub}{\subseteq}

\newcommand{\M}{\mathrm{M}}
\newcommand{\GL}{\mathrm{GL}}
\newcommand{\Ad}{\mathrm{Ad}}
\newcommand{\Inn}{\mathrm{Int}}

\newcommand{\diag}{\mathrm{diag}}

\newcommand{\ket}[1]{\vert\mathit{#1}\rangle}

\newcommand{\SL}{\mathrm{SL}}
\newcommand{\Sp}{\mathrm{Sp}}

\newcommand{\U}{\mathrm{U}}
\renewcommand{\sl}{\mathrm{sl}}
\renewcommand{\P}{\mathcal{P}}
\renewcommand{\H}{\mathcal{H}}

\newcommand{\be}{\begin{equation}}
\newcommand{\ee}{\end{equation}}
\newcommand{\bed}{\begin{displaymath}}
\newcommand{\eed}{\end{displaymath}}

\begin{document}

\title{On Clifford groups in quantum computing}

\author{J Tolar $^1$}
\address{
 $^1$ Department of Physics \\
 Faculty of Nuclear Sciences and  Physical  Engineering\\
 Czech Technical University in Prague\\ B\v rehov\'a 7,
 115 19 Prague 1, Czech Republic}
\eads{\mailto{jiri.tolar@fjfi.cvut.cz}}

\begin{abstract}
The term Clifford group was introduced in 1998 by D. Gottesmann in
his investigation of quantum error-correcting codes. The simplest
Clifford group in multiqubit quantum computation is generated by a
restricted set of unitary Clifford gates - the Hadamard,
$\pi/4$-phase and controlled-X gates. Because of this restriction
the Clifford model of quantum computation can be efficiently
simulated on a classical computer (the Gottesmann-Knill theorem).
However, this fact does not diminish the importance of the Clifford
model, since it may serve as a suitable starting point for a
full-fledged quantum computation.

In the general case of a single or composite quantum system with
finite-dimensional Hilbert space the finite Weyl-Heisenberg group of
unitary operators defines the quantum kinematics and the states of
the quantum register. Then the corresponding Clifford group is
defined as the group of unitary operators leaving the
Weyl-Heisenberg group invariant. The aim of this contribution is to
show that our comprehensive results on symmetries of the Pauli
gradings of quantum operator algebras -- covering any single as well
as composite finite quantum systems -- directly correspond to
Clifford groups defined as quotients with respect to $\U(1)$.
\end{abstract}

 \section{Introduction}

In quantum mechanics of single $N$-level systems in Hilbert spaces
of finite dimension $N$, the basic operators are the
\textit{generalized Pauli matrices}. They generate the finite
\textit{Weyl-Heisenberg group} (defining quantum kinematics and the
states of the quantum register) as a subgroup of the unitary group
$\U(N)$  \cite{Weyl,Schwinger,StovTolar84,Vourdas}.

Its normalizer within the unitary group $\U(N)$, or in other words
the largest subgroup of the unitary group having the Weyl-Heisenberg
group as a normal subgroup, is in the papers on quantum information
conventionally called the \textit{Clifford group} \cite{Gottesman}.
Since this normalizer necessarily contains the continuous group
$\U(1)$ of phase factors, some authors adopt an alternative
definition of the Clifford group as the quotient of the normalizer
with respect to $\U(1)$ \cite{Koenig}. In this paper we call it the
\textit{Clifford quotient group}.

Our results on symmetries of the Pauli gradings of quantum operator
algebras \cite{HPPT02,KorbTolar10,KorbTolar12} account for all
possible Clifford quotient groups corresponding to arbitrary single
or composite quantum systems. This contribution presents a brief
review of their description. In Sect. 2 the Weyl-Heisenberg groups
of $N$-level quantum systems are defined as subgroups of $\U(N)$ for
$N=2,3,\dots$. In Sect. 3 the corresponding Clifford quotient groups
are described and in Sect. 4 the Clifford quotient groups for
arbitrary composite quantum systems are shortly introduced.

\section{Weyl-Heisenberg groups of single $N$-level systems}

In finite-dimensional quantum mechanics of a single $N$-level system
the $N$-dimensional Hilbert space $\mathcal{H}_{N}=\mathbb{C}^N$ has
an orthonormal basis $\mathcal{B} = \left\lbrace\ket{0}, \ket{1},
\ldots \ket{N-1}\right\rbrace$. The basic unitary operators $Q_N$,
$P_N$ are defined by their action on the basis \cite{Weyl,Schwinger}
    \begin{eqnarray}
     Q_N \ket{j}&=& \omega_N^j \ket{j}, \\
     P_N \ket{j} &=& \ket{j-1 \pmod{N}},
    \end{eqnarray}
where $j=0,1,\ldots,N-1$, $\omega_N = \exp(2\pi i/N)$. This is the
well-known \textit{clock-and-shift representation} of the basic
operators $Q_N$, $P_N$. In the canonical or computational basis
$\mathcal{B}$ the operators $Q_N$ and $P_N$ are represented by the
\textit{generalized Pauli matrices}
    \begin{equation}
 Q_N = \mbox{diag}\left(1,\omega_N,\omega_N^2,\cdots,\omega_N^{N-1}\right)
    \end{equation}
and
    \begin{equation}
     P_N = \left(
    \begin{array}{cccccc}
     0 & 1 & 0& \cdots & 0 & 0 \\
     0 & 0& 1&  \cdots & 0 & 0 \\
     0 & 0 & 0&\cdots & 0 & 0 \\
     \vdots &   & & \ddots &   & \\
     0 & 0 &0 & \cdots & 0 & 1 \\
     1 & 0 &0 &\cdots & 0 & 0
    \end{array} \right) .
    \end{equation}
Their commutation relation
\begin{equation}
    P_N Q_N = \omega_N Q_N P_N
     \end{equation}
expresses the minimal non-commutativity of the operators $Q_N$ and
$P_N$. Further, they are of the order $N$,
 $$ P_N^N = Q_N^N = I, \quad \omega_N^N = 1.$$
The \textit{Pauli group} $\Pi_N$ of order $N^3$ is generated by
$\omega_N$, $Q_N$ and $P_N$ \cite{HPPT02}
     \begin{equation}
       \Pi_N = \left\lbrace \omega_N^l Q^i_N P_N^j |
     l,i,j = 0,1,2,\ldots,N-1\right\rbrace.
     \end{equation}
Elements of $\mathbb{Z}_N = \left\lbrace 0,1,\ldots N-1
 \right\rbrace $ label the vectors of the canonical basis
$\mathcal{B}$ with the physical interpretation that $\ket{j}$ is the
(normalized) eigenvector of position at $j \in \mathbb{Z}_N$. In
this sense the cyclic group $\mathbb{Z}_N$ plays the role of the
\textit{configuration space for an $N$-level quantum system}.

Now the Clifford group to be constructed should contain the
corresponding \textit{Weyl-Heisenberg group} as its normal subgroup.
The phase factors emerging in the Clifford group lead to the
necessity to define the Weyl-Heisenberg groups $H(N)$ in even
dimensions $N$ by doubling the Pauli group so that $H(N)$ for even
$N$ contains the Pauli group $\Pi_N$ as its subgroup. For this
purpose the phase factor $ \tau_N = - e^{\frac{\pi i}{N}} $ is
introduced such that $\tau_N^2 =\omega_N$
\cite{Appleby11,Appleby04,Hostens} (see also
\cite{BengtssonZyczkowski}). Then the Weyl-Heisenberg groups
   $$  H(N) = \Pi_N = \left\lbrace \omega_N^l Q^i_N P_N^j  \ |
\     l,i,j = 0,1,\ldots,N-1\right\rbrace \quad \mbox{for odd} \ N
,$$
$$  H(N) = \left\lbrace \tau_N^l Q^i_N P_N^j \ |
\     i,j = 0,1,\ldots,N-1, l = 0,1,\ldots,2N-1 \right\rbrace
     \quad \mbox{for even} \  N, $$
are of orders $N^3$, $2N^3$, respectively. Note that for $N =2$ we
have $H(2) =  \langle \tau_2 I_2, Q_2, P_2 \rangle $, where $\tau_2
= -i$, $Q_2 = \sigma_z $, $P_2 = \sigma_x $; $H(2)$ is  of order 16.

The \textit{center $Z(H(N))$ of the Weyl-Heisenberg group} is the
set of all those elements of $H(N)$ which commute with all elements
in $H(N)$.
For odd $N$ we have
$$Z(H(N)) = \{ \omega_N, \omega_N^2, \cdots, \omega_N^N = 1 \} = \{
\tau_N, \tau_N^2, \cdots, \tau_N^N = 1 \},$$
 while for even $N$
$$Z(H(N)) = \{\tau_N, \tau_N^2, \cdots, \tau_N^{2N} = 1 \}.
 $$
Since the center is a normal subgroup, one can go over to the
quotient group $$\P_N =H(N) / Z(H(N)).$$ Its elements are the cosets
labeled by pairs of exponents $(i,j)$, $i,j = 0,1,\ldots,N-1$. The
quotient group $\P_N$ is usually identified with the \textit{finite
phase space} $\mathbb{Z}_N \times \mathbb{Z}_N$. Namely, denoting
the cosets corresponding to elements $(i,j)$ of the phase space by
$Q^i P^j= \left\lbrace \tau_N^l Q^i_N P_N^j \right\rbrace$, the
correspondence
    \begin{equation}
\Phi : H(N) / Z(H(N)) \rightarrow \mathbb{Z}_N \times \mathbb{Z}_N :
Q^i P^j \mapsto (i,j),
     \end{equation}
is an \textit{isomorphism of Abelian groups}, since
   \begin{equation}
     \Phi\left(\left(Q^i P^j\right)\left(Q^{i'} P^{j'}\right)\right) =
\Phi\left(\left(Q^i P^j\right)\right)\Phi\left(\left(Q^{i'}
P^{j'}\right)\right)
  = (i,j) + (i',j') = (i+i',j+j').
     \end{equation}

\section{Clifford groups of single $N$-level quantum systems}

As already mentioned, in quantum information the term
\textit{Clifford group} means the group of symmetries of the
Weyl-Heisenberg group in the following sense
\cite{Appleby11,Appleby04}:

 \textbf{Definition}\\
\textit{The Clifford group comprises all unitary operators $X \in
\U(N)$ for which the $\Ad$-action preserves the subgroup $H(N)$ in
$\U(N)$,
i.e. the Clifford group is the normalizer $N_{\U(N)}(H(N))$.}\\

In this sense the Clifford group consists of all those $X \in \U(N)$
such that their $\Ad$-action leaves $H(N)$ invariant,
 $$ \Ad_X H(N) = XH(N) X^{-1} = H(N).$$
But $H(N)$ is generated by $\tau_N$, $Q_N$ and $P_N$, so the
Clifford group consists of all $X \in \U(N)$ such that
 $$
 \Ad_X Q_N = XQ_N X^{-1} \in H(N) \quad \mathrm{and} \quad
  \Ad_X P_N = XP_N X^{-1} \in H(N).
 $$

Some authors use the notion of \textit{Clifford operations}
\cite{Hostens}. They are the one-step unitary evolution operators
acting on the nodes of a quantum register and are physically
realized by so-called \textit{Clifford gates}.\footnote{Note that
the original name Clifford group is connected with Clifford
algebras. For instance, the Clifford algebra generated by four Dirac
matrices $\gamma^{\mu}$ is the real Dirac algebra $D$ of dimension
$2^4 = 16$. The corresponding Clifford group is the universal
covering of the Lorentz group $O(3,1)$ of isometries of the real
vector space $\mathbb{R}^4$ with an indefinite inner product,
spanned by $\gamma^{0}$, $\gamma^{1}$, $\gamma^{2}$ and $\gamma^{3}$
such that
$$ S \gamma^{\mu} S^{-1} = \Lambda^{\mu}_{\nu}\gamma^{\nu}.$$ The
real Pauli algebra is the even part of $D$ of dimension $2^3 = 8$
and its isometries form the universal covering of the orthogonal
group $O(3)$. }

Since the subgroup $H(N)$ is a normal subgroup of the normalizer,
one can formally write a short exact sequence of group homomorphisms
$$      1 \rightarrow H(N) \rightarrow N_{\U(N)}(H(N))
\rightarrow N_{\U(N)}(H(N))/H(N)  \rightarrow 1 .
  $$
It turns out that the full structure of the normalizer
$N_{U(N)}(H(N))$ for arbitrary $N$ is complicated by phase factors
and rather difficult to describe in general \cite{Hostens}. In order
to get insight into the structure of the normalizer \textit{up to
arbitrary phase factors} we turn to the definition of the
\textit{Clifford quotient group} $C(N)$ as the quotient
\cite{Koenig}
$$  C(N)=N_{U(N)}(H(N))/ U(1). $$
Its elements are the cosets $\{e^{i\alpha} X  \}$. The following
lemma is crucial for our alternative view of the Clifford quotient
group.

 \textbf{Lemma}\\
 \textit{Let $X,Y \in \U(N)$. Then the equality
$ \Ad_X A = \Ad_Y A  \quad \mbox{holds for all} \quad A\in H(N)
 $ if and only if   $ X=e^{i\alpha}Y$.}
\\
The proof of the converse implication is trivial. For the proof of
the direct implication one writes down the first identity in the
form $Y^{-1}XA=AY^{-1}X$, i.e. $Y^{-1}X$ commutes with all $A\in
H(N)$. But the elements of $H(N)$ form an irreducible set, hence by
Schur's lemma $Y^{-1}X$ is proportional to the unit matrix,
$Y^{-1}X=e^{i\alpha}I_N$.$\Box$\\

In accordance with the Lemma, instead of the cosets $\{e^{i\alpha} X
\}$ one can equivalently consider $\Ad$-actions induced by the
elements $X$ of the normalizer. Generally, the mappings
 $ \Ad_X \ : \ A \rightarrow XAX^{-1}$, where $A\in \GL(N,\C)$, are
inner automorphisms of $\GL(N,\C)$ induced by elements $X\in
\GL(N,\C)$. In our case we need the subgroup $\mathcal{M_N}$ of
$\Inn (\GL(N,\C))$ generated by unitary operators, $\mathcal{M_N} =
\{ \Ad_X \vert X\in \U(N) \}$. In fact, if inner automorphisms
$\Ad_X$ transform unitary operators $A$ in unitary operators
$A'=XAX^{-1}$, then to each $X\in \GL(N,\C)$ there exists a unitary
operator $U\in \U(N)$ such that $\Ad_U =  \Ad_X$ and $U$ is unique
up to a phase factor.

In this equivalent approach the Clifford quotient group $C(N)$ can
be studied as a subgroup of $\mathcal{M_N}$ -- the subgroup of those
$\Ad$-actions which preserve the Weyl-Heisenberg group $H(N)$. But
$C(N)$, consisting of the cosets $\{e^{i\alpha} X \}$ leaving $H(N)$
invariant, contains the cosets $\{e^{i\alpha} A \}$ of operators
$A\in H(N)$ as a subgroup, and this subgroup is isomorphic to the
group of $\Ad$-actions $\Ad_A$, $A\in H(N)$. Note that $\Ad$-actions
of unitary operators commute $\Ad_A \Ad_B = \Ad_B \Ad_A$ if and only
if there exists $q \in \C^*$ such that $AB = q BA$ and $q^N = 1$.
Hence in this picture the group of $\Ad$-actions $\Ad_A$, $A\in
H(N)$, is generated by the commuting $\Ad$-actions $\Ad_{Q_N}$ and
$\Ad_{P_N}$,
 $$  \{ \Ad_{Q_N^i
P_N^j} \vert i,j = 0,1,\dots,N-1 \} \cong \mathbb{Z}_N \times
\mathbb{Z}_N \cong \P_N . $$

\textbf{Proposition}\\
\textit{The Clifford quotient group $C(N)$ is isomorphic to the
subgroup of those inner automorphisms in $\mathcal{M_N}$ which
preserve $\P_N$, i.e. $C(N)$ is the normalizer of $\P_N$ in
$\mathcal{M_N}$,
$$C(N) \cong N_{\mathcal{M_N}}(\P_N).$$}

We will show now that the short exact sequence of homomorphisms of
subgroups of $\mathcal{M_N}$
 \be  \label{9}      1 \rightarrow \P_N
\rightarrow N_{\mathcal{M_N}}(\P_N)
 \rightarrow N_{\mathcal{M_N}}(\P_N)/\P_N  \rightarrow 1  \ee
can be fully decoded \cite{HPPT02}.

Obviously, $\Ad$-actions of elements of $\P_N$ leave $\P_N$
invariant. Then the elements of the quotient group
$N_{\mathcal{M}}(\P_N)/\P_N$ are the cosets corresponding to
possibly non-trivial transformations of $\P_N$ forming a symmetry
(or Weyl) group. Let us consider the $\Ad$-actions  $\Ad_X (A) = X A
X^{-1}$, where $X \in U(N)$, on elements $A \in H(N)$, which induce
permutations of cosets in $H(N) / Z(H(N)$. We consider them to be
equivalent if, for each pair $(i,j) \in \mathbb{Z}_N \times
\mathbb{Z}_N$, they define the same transformation of cosets in
$H(N) / Z(H(N))$:
$$
\Ad_Y \sim \Ad_X \quad \Leftrightarrow \quad Y Q^i P^j Y^{-1}= X Q^i
P^j X^{-1}.
  $$
We have seen that the group $\P_N$ has two generators, $\Ad_{Q_N}$
and $\Ad_{P_N}$, corresponding to cosets $Q$ and $P$. Hence if
$\Ad_Y$ induces a permutation of elements in $\P_N$, then there must
exist $a,b,c,d \in \mathbb{Z}_N$ such that
 $$ Y Q Y^{-1} = Q^a P^b \quad \mbox{and} \quad Y P Y^{-1} = Q^c P^d.
 $$

It follows that to each equivalence class of $\Ad$-actions $\Ad_Y$ a
quadruple $(a,b,c,d)$ of elements in $\mathbb{Z}_N$ is assigned. In
matrix notation
$$
 \Ad_Y \
\left( \begin{array}{c}
        1 \\
        0
       \end{array}\right) =
\left( \begin{array}{c}
        a \\
        b
       \end{array}\right) =  \left( \begin{array}{cc}
        a & c \\
        b & d
       \end{array}\right)
\left( \begin{array}{c}
        1 \\
        0
       \end{array}\right),
$$
$$
\Ad_Y \ \left( \begin{array}{c}
        0 \\
        1
       \end{array}\right)
 = \left( \begin{array}{c}
        c \\
        d
       \end{array}\right) =  \left( \begin{array}{cc}
        a & c \\
        b & d
       \end{array}\right)
\left( \begin{array}{c}
        0 \\
        1
       \end{array}\right).
$$
Now inserting the relations
$$
    \Ad_Y Q_N = \mu Q_N^a P_N^b,  \qquad   \Ad_Y P_N = \nu Q_N^c P_N^d,
  $$
into the basic commutation condition
$$
  \Ad_Y (P_N Q_N) = \omega_N \Ad_Y (Q_N P_N),
$$
we find
$$
 \omega_N^{ad-1} = \omega_N^{bc} \quad \mbox{i.e.} \quad
ad - bc =\det \left( \begin{array}{cc}
        a & c \\
        b & d
       \end{array}\right)=1 \pmod N.
$$

This result can also be stated as the value in $\mathbb{Z}_N$ of a
bilinear alternating non-degenerate form (symplectic form) on the
finite phase space $\P_N = \mathbb{Z}_N \times \mathbb{Z}_N$.
Namely, the symplectic form is the bilinear mapping
 $ q \ : \ \P_N \times \P_N \longrightarrow  \Z_N  $, where
$$
q \ : \
 ((i,j)(i',j')) \mapsto  i'j - ij' = (i,j)\left( \begin{array}{cc}
        0 & -1 \\
        1 & 0
       \end{array}\right)
\left( \begin{array}{c}
        i' \\
        j'
       \end{array}\right)
= \det \left( \begin{array}{cc}
        i' & j' \\
        i & j
       \end{array}\right).
$$

\textbf{Theorem} \cite{HPPT02}\\
 \textit{For integer $N\geq 2$ there
is an isomorphism $\Phi$ between the set of equivalence classes of
$\Ad-$actions $\Ad_Y$ which induce permutations of cosets, and the
group $\SL(2,\mathbb{Z}_N)$ of $2 \times 2$ matrices with
determinant equal to $1 \, \pmod{N}$,
       \begin{equation}
        \Phi(\Ad_Y) = \left( \begin{array}{cc}
        a & c \\
        b & d
       \end{array}\right), \qquad a,b,c,d \in \mathbb{Z}_N.
       \end{equation}
The action of these automorphisms on $\P_N$ is given by (left)
action of $\SL(2,\mathbb{Z}_N)$ on elements $(i,j)^T$ of the phase
space $\P_N=\mathbb{Z}_N \times \mathbb{Z}_N$,
       \begin{equation}
       \Ad_Y \
 \left( \begin{array}{c}
        i \\
        j
       \end{array}\right)
= \left( \begin{array}{c}
        i' \\
        j'
       \end{array}\right)
=   \left( \begin{array}{cc}
        a & c \\
        b & d
       \end{array}  \right)
\left( \begin{array}{c}
        i \\
        j
       \end{array}\right).
       \end{equation} }

It follows from the above results that the groups entering (\ref{9})
are isomorphic to
\begin{eqnarray}
   \P_N & \cong &  \mathbb{Z}_N \times \mathbb{Z}_N,\\
N_{\mathcal{M_N}}(\P_N)/\P_N  & \cong & \SL(2,\mathbb{Z}_N),\\
C(N) = N_{\mathcal{M_N}}(\P_N) & \cong & (\mathbb{Z}_N \times
\mathbb{Z}_N) \rtimes \SL(2,\mathbb{Z}_N),
\end{eqnarray}
where $\rtimes$ denotes a semidirect product. Here $\P_N$ is a
normal subgroup of the normalizer $N_{\mathcal{M_N}}(\P_N)$ with two
generators $\Ad_{Q_N}$, $\Ad_{P_N}$.

Summarizing, the \textit{Clifford quotient group} $C(N)$ is
isomorphic to the normalizer of the Abelian subgroup $\P_N$ in the
group of unitary inner automorphisms $\mathcal{M_N}$. Since it
contains all inner automorphisms transforming the phase space into
itself, it necessarily contains $\P_N$ as an Abelian semidirect
factor. The symmetry (or Weyl) group is then isomorphic to the
quotient group of the normalizer with respect to $\P_N$.

The generators of the normalizer $N_{\mathcal{M_N}}(\P_N)$ are
$\Ad_{Q_N}$, $\Ad_{P_N}$ and $\Ad_{S_N}$, $\Ad_{D_N}$ defined below
\cite{HPPT02}: The unitary Sylvester matrix $S_N$ is the matrix of
the discrete Fourier transformation (for $N=2$ the Hadamard gate):
   \begin{equation}
     (S_{N})_jk = \frac{\omega_N^{jk}}{\sqrt{N}}.
   \end{equation}
It acts on $Q_N$ and $P_N$ according to
   \begin{equation}
   S_N Q_N S_N^{-1}= P_N^{-1}  \quad  S_N P_N S_N^{-1} = Q_N ,
   \end{equation}
The unitary operator $D_N$  (for $N=2$ the phase gate) is diagonal,
$$
D_N = \mathrm{diag} \ (d_0,d_1,\dots,d_{N-1}),
$$
where $d_j = \tau_N^{j(1-j)}$ if $N$ is odd, $d_j = \tau_N^{j(N-j)}$
if $N$ is even. It acts on $Q_N$ and $P_N$ according to
$$
  D_{N}Q_N D_N^{-1} = Q_N, \qquad
  D_{N}P_N D_N^{-1} = \alpha_N Q_N P_N \,
$$
where $\alpha_N = 1$ for $N$ odd and $\alpha_N = \tau_N^{N+1}$ for
$N$ even.

Given the prime decomposition of $N =
\prod\limits_{i=1}^{r}p_i^{k_i}$, the general formula for the number
of elements of $\SL(2,\mathbb{Z}_N)$ is the following multiplicative
function of number theory:
$$    |\SL(2,\mathbb{Z}_N)| =
   N^3 \prod\limits_{i=1}^{r}\left(1 - \frac{1}{p_i^2}\right).  $$
Some cardinalities are given in the table:
       \begin{center}
\renewcommand{\arraystretch}{1.9}
\begin{tabular}{|l|rrr|}
\hline
 $N$ & $|\P_N|$ & $|\SL(2,\mathbb{Z}_N)|$ & $|\P_N\rtimes \SL(2,\mathbb{Z}_N)|$\\
  \hline
 2  &    4   &     6     &   24    \\
 3  &    9    &    24    &   216   \\
 4  &  16     &     48   &   768      \\
 5   &  25    &    120   &  3000        \\
 6   &  36    &    144   &  5184        \\
 7   &  49    &    336    &  16464       \\
 8   &  64    &    384    &  24576       \\
\hline
\end{tabular}
\end{center}


\textbf{Example ~ $N=2$}  \cite{Sloane}:~ The phase space consists
of 4 elements $(0,0)$, $(1,0)$, $(0,1)$, $(1,1)$. The group
$\SL(2,\mathbb{Z}_2)$ has 6 elements
$$
\left( \begin{array}{cc} 1 & 0 \\  0 & 1 \end{array} \right), \left(
\begin{array}{cc} 0 & 1 \\  1 & 0 \end{array} \right), \left(
\begin{array}{cc} 1 & 0 \\  1 & 1 \end{array} \right), \left(
\begin{array}{cc} 1 & 1 \\  0 & 1 \end{array} \right), \left(
\begin{array}{cc} 1 & 1 \\  1 & 0 \end{array} \right), \left(
\begin{array}{cc} 0 & 1 \\  1 & 1 \end{array} \right),
$$
and acts transitively on the orbit $\left\lbrace
(1,0),(0,1),(1,1)\right\rbrace$. Unitary operators $S_2$ and $D_2$
are
$$
S_2 =\frac{1}{\sqrt{2}} \left( \begin{array}{cc} 1 & 1 \\ 1 & -1
\end{array} \right), \quad D_2 =\left( \begin{array}{cc} 1 & 0 \\  0
& -i \end{array} \right).
$$
The finite Clifford group generated by $S_2$ and $D_2$ has $24\times
8 =192$ elements, since $(S_2 D_2)^3 = \eta I_2$ is of order 8:
$$
\left( \begin{array}{cc} 1 & 0 \\  0 & \alpha \end{array} \right),
\left(
\begin{array}{cc} 0 & 1 \\  \alpha & 0 \end{array} \right),
\frac{1}{\sqrt{2}}\left(
\begin{array}{cc} 1 & \beta \\  \alpha & -\alpha\beta \end{array} \right),
\eta^{\nu}\left(
\begin{array}{cc} 1 & 0 \\  0 & 1 \end{array} \right),
$$
where $\eta=\exp(i\frac{\pi}{4})$, $\nu=0,1,\dots,7$, $\alpha,\beta
\in \{1,i,-1,-i\}$. Note that $Q_2=\sigma_3$, $P_2=\sigma_1$ as well
as their products and powers are generated by $S_2$ and $D_2$.

In this example the finite Clifford group is defined neither as the
whole normalizer of the Heisenberg-Weyl group nor as its quotient
with respect to $U(1)$, but a certain finite subgroup of unitary
operators in the normalizer \cite{BP17}.

In mathematical terms the motivation for this construction comes
from Cartan's method for simple Lie algebras. There in the first
step the maximal commutative subalgebra containing semi-simple
elements is identified. This Cartan subalgebra generates a
commutative subgroup (complex torus) in the corresponding complex
Lie group or the torus in the corresponding compact Lie group. Then
the eigenspaces of its $\Ad$-action are the root subspaces of the
Lie algebra. They are permuted by the Weyl group of symmetries of
the root subspaces.

The construction of the \textit{finite Clifford group} can then be
seen as an analogue of the \textit{Demazure-Tits finite group} for
simple Lie algebras \cite{MichelPateraSharp}. In dimension $N$ the
finite Clifford group is a special finite subgroup of the normalizer
$N_{U(N)}(H(N))$ completely defined by its generators $\tau_N$,
$Q_N$, $P_N$ and $S_N$, $D_N$. Moreover, it was shown in \cite{BP17}
that the finite Clifford group for odd $N$ is generated only by
$S_N$ and $D_N$.

\section{Clifford quotient groups of multipartite systems}

Our further results concern detailed description of groups of
symmetries of finite Weyl-Heisenberg groups for \textit{finitely
composed quantum systems consisting of subsystems with arbitrary
dimensions}. We have fully described these symmetries on the level
of $\Ad$-actions. In our notation \cite{KorbTolar12} the most
general \textit{symmetry (or Weyl) groups} in the sense of
\cite{PZ89} are $\Sp_{[n_{1},\dots,n_{k}]}$, where the indices
denote arbitrary dimensions of the constituent Hilbert spaces.

More in detail, let the Hilbert space of a composite system be the
tensor product
 $ \H_{n_1}\otimes\dots\otimes\H_{n_k}$
of dimension $N=n_{1}\dots n_k$, where $n_{1},\dots,n_{k}\in\N$. For
the \textit{composite system}, quantum phase space is the Abelian
subgroup of $\Inn(\GL(N,\mathbb{C}))$ defined by
$$\mathcal{P}_{(n_{1},\dots,n_{k})}=
\set{\Ad_{M_{1}\otimes\cdots\otimes M_{k}}}{M_{i}\in H(n_{i})}.$$

The \textit{Clifford quotient group}, or the normalizer of this
Abelian subgroup in the group of unitary inner automorphisms of
$\GL(N,\mathbb{C})$, contains all unitary inner automorphisms
transforming the phase space into itself, hence necessarily
\textit{contains $\mathcal{P}_{(n_{1},\dots,n_{k})}$ as an Abelian
semidirect factor}. The symmetry (or Weyl) group is then given by
the quotient group of the normalizer with respect to this Abelian
subgroup.

The generating elements of $\mathcal{P}_{(n_{1},\dots,n_{k})}$ are
the inner automorphisms
$$e_{j}:=\Ad_{A_{j}} \quad \textrm{for} \quad j=1,\dots,2k,$$
 where (for $i=1,\dots,k$)
$$A_{2i-1}:=I_{n_{1}\cdots n_{i-1}}\otimes P_{n_{i}}
\otimes I_{n_{i+1}\cdots n_{k}},\\
 A_{2i}:=I_{n_{1}\cdots n_{i-1}}\otimes Q_{n_{i}}
 \otimes I_{n_{i+1}\cdots n_{k}}.
 $$
The normalizer of $\mathcal{P}_{(n_{1},\dots,n_{k})}$ in
$\Inn(\GL(n_{1}\cdots n_{k},\C))$ will be denoted
$$\mathcal{N}(\mathcal{P}_{(n_{1},\dots,n_{k})}):=
N_{\Inn(\GL(n_{1}\cdots
n_{k},\C))}(\mathcal{P}_{(n_{1},\dots,n_{k})}),$$
 We need also the normalizer of $\P_{n}$ in $\Inn(\GL(n,\C))$,
$$\mathcal{N}(\P_{n}):=N_{\Inn(\GL(n,\C))}(\P_{n}),$$
 and
$$\mathcal{N}(\P_{n_{1}})\times\cdots\times\mathcal{N}(\P_{n_{k}}):=
\set{\Ad_{M_{1}\otimes \cdots\otimes M_{k}}}{M_{i}\in
\mathcal{N}(\P_{n_{i}})}\sub\Inn(\GL(N,\C)).$$
 Further,
$$\mathcal{N}(\P_{n_{1}})\times\cdots\times\mathcal{N}(\P_{n_{k}})
\subseteq\mathcal{N}(\mathcal{P}_{(n_{1},\dots,n_{k})}).
  $$

Now the \textit{symmetry group} $\Sp_{[n_{1},\dots,n_{k}]}$ is
defined in several steps. First let
$\mathcal{S}_{[n_{1},\dots,n_{k}]}$ be a set consisting of $k\times
k$ matrices $H$ of $2\times 2$ blocks
$$H_{ij}=\frac{n_{i}}{\gcd(n_{i},n_{j})}A_{ij}$$ where $2\times 2$
matrices $A_{ij}\in\M_{2}(\Z_{n_{i}})$ for $i,j=1,\dots,k$. Then
$\mathcal{S}_{[n_{1},\dots,n_{k}]}$ is (with the usual matrix
multiplication) a monoid. Next, for a matrix $H\in
\mathcal{S}_{[n_{1},\dots,n_{k}]}$, we define its adjoint
$H^{\ast}\in \mathcal{S}_{[n_{1},\dots,n_{k}]}$ by
$$(H^{\ast})_{ij}=\frac{n_{i}}{\gcd(n_{i},n_{j})}A_{ji}^{T}.$$
Further, we need a skew-symmetric matrix
$$J=\diag(J_{2},\dots,J_{2})\in\mathcal{S}_{[n_{1},\dots,n_{k}]}$$
where $$J_{2}=\left(\begin{array}{cc}0&1\\-1&0\end{array}\right).$$
Then  the \textit{symmetry group} is defined as
$$\Sp_{[n_{1},\dots,n_{k}]}:=\set{H\in
\mathcal{S}_{[n_{1},\dots,n_{k}]}}{\ H^{\ast}JH= J}$$ and is a
finite subgroup of the monoid $\mathcal{S}_{[n_{1},\dots,n_{k}]}$.

Our first theorem states the group isomorphism:

\textbf{Theorem 1} \cite{KorbTolar12}\\
$$\mathcal{N}(\mathcal{P}_{(n_{1},\dots,n_{k})})/
\mathcal{P}_{(n_{1},\dots,n_{k})}\cong \Sp_{[n_{1},\dots,n_{k}]}.$$

Our second theorem describes the generating elements of the
normalizer:

\textbf{Theorem 2} \cite{KorbTolar12}\\
\textit{The normalizer
$\mathcal{N}(\mathcal{P}_{(n_{1},\dots,n_{k})})$ is generated by
$$\mathcal{N}(\P_{n_{1}})\times\cdots\times\mathcal{N}(\P_{n_{k}})
 \quad \textrm{and} \quad  \{\Ad_{R_{ij}}\},$$
  where (for $1\leq i<j\leq k$)
$$R_{ij}=I_{n_{1}\cdots n_{i-1}}\otimes\mathrm{diag}
(I_{n_{i+1}\cdots n_{j}},T_{ij},\dots,T_{ij}^{n_{i}-1}) \otimes
I_{n_{j+1}\cdots n_{k}}$$
and
$$T_{ij}=I_{n_{i+1}\cdots
n_{j-1}}\otimes Q_{n_{j}}^{\frac{n_{j}}{\gcd(n_{i},n_{j})}}.$$}

An important special case is

\textbf{Corollary} \cite{KorbTolar12}\\
\textit{If $n_1=\dots=n_k= n$, i.e. $N=n^k$, the symmetry group is
$\Sp_{[n,\dots,n]}\cong \Sp_{2k}(\Z_n)$.}\\

These cases are of particular interest, since they uncover
\textit{symplectic symmetry} of $k$-partite systems composed of
subsystems with the same dimensions. This circumstance was found, to
our knowledge, first by \cite{PST06} for $k=2$ under additional
assumption that $n=p$ is prime, leading to $\Sp(4,\mathrm{F}_p)$
over the field $\mathrm{F}_p$. We have generalized this result
\cite{KorbTolar10} to bipartite systems with arbitrary $n$
(non-prime) leading to $\Sp(4,\mathbb{Z}_n)$ over the modular ring,
and also to multipartite systems \cite{KorbTolar12}. The
corresponding result was independently obtained by \cite{Han10} who
studied symmetries of the tensored Pauli grading of the Lie algebra
$\sl(n^k,\C)$ (see also \cite{VourdasBanderier}).

\section{Illustrative examples}

Consider a \textit{bipartite system} created by coupling two single
multi-level subsystems with arbitrary dimensions $n$, $m$, i.e.
 $$G=\Z_n \times \Z_m \quad \mbox{and} \quad \H = \H_n
\otimes \H_m.$$
 The corresponding finite Weyl-Heisenberg group is
embedded in $\U(N)$, $N=nm$. Via \textit{inner automorphisms} it
induces an Abelian subgroup in $\Inn(\GL(N,\mathbb{C}))$.

The \textit{Clifford quotient group}, or the normalizer of this
Abelian subgroup in the group of inner automorphisms of
$\GL(N,\mathbb{C})$, contains all unitary inner automorphisms
transforming the phase space into itself, hence necessarily
\textit{contains $\P_{(n,m)}$ as an Abelian semidirect factor}. The
symmetry is then given by the quotient group of the normalizer with
respect to this Abelian subgroup.

According to the Corollary the case $n=m$, $N=n^2$, corresponds to
the symmetry group
 $$\Sp_{[n,n]}\cong \Sp(4,\Z_n).$$
If $N=nm$, $n,m$ coprime, the symmetry group is
 $$\Sp_{[n,m]} \cong \SL(2,\Z_n)\times \SL(2,\Z_m)\cong \SL(2,\Z_{nm}).$$
Further, if $d=gcd(n,m)$, $n=ad$, $m=bd$, the finite configuration
space can be further decomposed under the condition that $a,b$ are
both coprime to $d$,
 $$
 G= \Z_{n}\times Z_{m} = Z_{ad}\times Z_{bd} \cong
 (Z_{a}\times Z_{d})\times (Z_{b}\times Z_{d}).
 $$
Thus the symmetry group is reduced to the direct product
 $$
 \Sp_{[n,m]} \cong \Sp_{[a,b]}\times \Sp(4,\Z_d)
 $$
For instance, if $n=15$, $m=12$, then $d=3$ is coprime to both $a=5$
and $b=4$, and also $a$ and $b$ are coprime, hence the symmetry
group is reduced to the standard types $\SL$ and $\Sp$,
 $$
 \Sp_{[n,m]}\cong \SL(2,\Z_a)\times \SL(2,\Z_b)\times \Sp(4,\Z_d).
 $$

Consider a still more general situation, for instance let $n=180$
and $m=150$. Then $d=30$, $a=180/30=6$ and $b=150/30=5$, hence $a$
divides $d$ and also $b$ divides $d$. In this case the reduction to
solely standard groups $\Sp$ and $\SL$ is not possible. One has to
break down the composite system consisting of two single systems
into its \textit{elementary building blocks} \cite{TolJPCS14}. We
decompose each of the finite configuration spaces
 $$
 \Z_{180}\times Z_{150} =(\Z_{2^2}\times Z_{3^2}\times  Z_{5})\times
                (\Z_{2}\times Z_{3}\times  Z_{5^2}),
 $$
and take notice of coprime factors $2.2^2$, $3.3^2$ and $5.5^2$
leading to the factorization of the symmetry group in agreement with
the elementary divisor decomposition
 $$
 \Sp_{[180,150]}\cong \Sp_{[2,2^2]}\times \Sp_{[3,3^2]}\times
 \Sp_{[5,5^2]}.
 $$

Thus if attention is paid to the elementary building blocks of
finite quantum systems (quantal degrees of freedom)
\cite{TolJPCS14}, then the symmetries can be reduced -- in
accordance with the elementary divisor decomposition -- to direct
products of finite groups of the types $\SL(2,\Z_n)$,
$\Sp(2k,\Z_n)$, and $\Sp_{[p^k,p^l,\dots]}$. The last type of
symmetry groups corresponds to a \textit{new class of Clifford
quotient groups of composite systems}. These symmetry groups like
$\Sp_{[p^k,p^l]}$ with indices given by different powers of the same
prime $p$ deserve to be added to the standard types $\Sp$ and $\SL$.

\section{Conclusion}


We have considered three definitions of the Clifford group. As the
most interesting seems not the first one -- the whole normalizer of
the Heisenberg-Weyl group, but the second -- its quotient with
respect to $U(1)$ or even the third -- a certain finite subgroup of
unitary operators in the normalizer like that presented in the
example in dimension $N=2$.

\section*{Acknowledgements}
The author is grateful to M. Havl\'{i}\v{c}ek, M. Korbel\'a\v{r} and
P. Novotn\'y for fruitful discussions on the subject of the paper.
Partial support of the Ministry of Education of Czech Republic (from
the research plan RVO:68407700) is gratefully acknowledged.

\end{document}